\documentclass[10pt, conference, compsocconf]{IEEEtran}
\ifCLASSINFOpdf
\else
\fi
\usepackage{amsmath}
\usepackage{url}

\usepackage{amssymb}
\usepackage{amstext}
\usepackage{amsfonts}
\usepackage{dsfont} 
\usepackage{stmaryrd}
\usepackage{xspace}
\usepackage{multirow}
\usepackage{graphicx}
\usepackage{array}
\usepackage{color}
\usepackage{subfigure}

\usepackage{cite}
\newtheorem{theorem}{Theorem}

\newtheorem{proof}{Proof}

\newtheorem{definition}{Definition}

\usepackage[english]{varioref}

\begin{document}

\title{A Cryptographic Approach for Steganography}



\author{\IEEEauthorblockN{Jacques M. Bahi, Christophe Guyeux, and Pierre-Cyrille Heam*\\}
\IEEEauthorblockA{FEMTO-ST Institute, UMR 6174 CNRS\\
  Computer Science Laboratory DISC\\
  University of Franche-Comt\'{e}, France\\
  \{jacques.bahi, christophe.guyeux, pierre-cyrille.heam\}@femto-st.fr}
*Authors are cited in alphabetic order
}

\maketitle

\begin{abstract}
In this research work, security concepts are formalized in steganography, 
and the common paradigms based on information theory are replaced by another ones 
inspired from cryptography, more practicable are closer than what is usually
done in other cryptographic domains. These preliminaries lead to a first proof
of a cryptographically secure information hiding scheme. 
\end{abstract}


\section{Introduction}

The usual manner for preserving privacy when communicating over public channels 
is by using cryptographic tools. 
Users cipher the data and send them over possibly insecure networks. 
Even if a third party intercepts these data, he or she will not understand them 
without having the secret key for deciphering.
In that well investigated scenario, anyone knows that a private message is 
transmitted through the public channel, but only authorized individuals (\emph{i.e.},
owners of the secret key) can understand it. 

A second approach investigated over two decades~\cite{Cachin2004}, and usually referred as
information hiding or steganography~\cite{bcg11b:ip,fgb11:ip}, aims at inserting a secret message into an 
innocent cover, in such a way that observers cannot detect the existence of
this hidden channel (for instance, images sent through the Internet). 
The goal in this field is to appear as innocent as possible:
observers should not think that something goes wrong with this public channel.
It must not cross their mind that sometimes the public channel is used
to transmit hidden messages. In that context, an attack is succeeded when 
the sleazy character of the channel is detected.
Tools used in that field are mainly based on artificial intelligence. They 
are called steganalyzers, and their main objective is to detect whether a 
given communication channel is possibly steganographied, or if it only
contains ``natural'' images. In case of detection, the unique countermeasure 
proposed by the literature is to stop the sleazy communication by closing the channel.
To sum up, the steganography community currently only focuses on the ability to detect
hidden channels, without investigating the consequences of this detection~\cite{gfb10:ip,bfg12:ip}.

However, observers have not necessarily the ability or the desire to stop the communication.
For instance, who can switch off the Internet? Furthermore, by stopping
the faked channel, attackers miss the opportunity to obtain more information about
the secret message and the intended receiver.
Finally, if attackers observe the communication, man can reasonably think
that they already knew in advance that this channel is sleazy (if not, why
they observe it?). The use of a steganalyzer on a channel only appears in the best 
situation as a reinforcement of their doubts or fears. In most operational
contexts, only sleazy channels are observed, and the questions are 
finally to determine~\cite{bg10b:ip}: (1) when the hidden messages have been
transmitted in this channel (among all the possibly faked images, how to determine 
the ones that really contain hidden information?), (2) what was the content of this message, 
and perhaps (3) who was the receiver among the observers.
These questions make sense only within a stegranographic context, that is,
when the channel is not ciphered. However, these important questionings have
never been regarded by the information hiding community.

In this paper, authors provide a cryptographic theoretical framework to study
this scenario related to steganography. Concrete illustrative examples of this framework 
of study are given thereafter. A first toy example is the hypothetical case of a
dissident blogger in a totalitarian state, who posts regularly and publicly 
information in his or her blog, while being severely watched by the
authorities. This blogger wants to transmit one day a secret message or a signal to 
an observer into confidence, without sounding the alarm in the authorities side.
Another example is an individual who is invigilated, because he is correctly suspected 
to be a spy. This agent cannot be arrested on a simple presumption, or on
the claim that the images he sent in his emails look sleazy.
Despite this surveillance, this spy wants to transmit one day a message
to his sponsor. The observers want to determine if an hidden message is 
really transmitted or not, to have a proof of such a transmission, together with
the content of the message, the date of transmission, and the 
targeted receiver if possible. 
Obviously, these situations are related to both cryptography
and steganography, however there is currently a lack of tool
allowing their study.
The key idea of this research work is to propose
algorithms such that observers cannot switch from doubts (sleazy channels) 
to certainties or proofs.


The remainder of this article is organized as follows. 
In Section~\ref{sec2}, generalities from steganography are discussed. 
The key concepts and main results are  presented in Section~\ref{sec4}.
Finally, Section~\ref{sec7} concludes 
this research work and details further investigations.

\section{Notions and Terminologies in Information Hiding}
\label{sec2}
In the following some common notions in the field of information hiding are recalled. 
We refer to \cite{STEG} for a complete survey of this subject.

\subsection{Information Hiding Security}

Robustness and security are two major concerns in information hiding.
These two concerns have been defined in \cite{Kalker2001} as follows. 
``Robust watermarking is a mechanism to create a communication channel that is multiplexed into original content [...]. It is required that, firstly, the perceptual degradation of the marked content [...] is minimal and, secondly, that the capacity of the watermark channel degrades as a smooth function of the degradation of the marked content. [...]. Watermarking security refers to the inability by unauthorized users to have access to the raw watermarking channel [...] to remove, detect and estimate, write or modify the raw watermarking bits.''

In the framework of watermarking and steganography, security has seen several important developments since the last decade~\cite{BarniBF03,Cayre2005,Ker06}. 
The first fundamental work in security was made by Cachin in the context of steganography~\cite{Cachin2004}. 
Cachin interprets the attempts of an attacker to distinguish between an innocent image and a stego-content as a hypothesis testing problem. 
In this document, the basic properties of a stegosystem are defined using the notions of entropy, mutual information, and relative entropy. 
Mittelholzer, inspired by the work of Cachin, proposed the first theoretical framework for analyzing the security of a watermarking scheme~\cite{Mittelholzer99}.

These efforts to bring a theoretical framework for security in steganography and watermarking have been followed up by Kalker, who tries to clarify the concepts (robustness \emph{vs.} security), and the classifications of watermarking attacks~\cite{Kalker2001}. 
This work has been deepened by Furon \emph{et al.}, who have translated Kerckhoffs' principle (Alice and Bob shall only rely on some previously shared secret for privacy), from cryptography to data hiding~\cite{Furon2002}. 
They used Diffie and Hellman methodology, and Shannon's cryptographic framework~\cite{Shannon49}, to classify the watermarking attacks into categories, according to the type of information Eve has access to~\cite{Cayre2005,Perez06}, namely: Watermarked Only Attack (WOA), Known Message Attack (KMA), Known Original Attack (KOA), and Constant-Message Attack (CMA).
Levels of security have been recently defined in these setups.
The highest level of security in WOA is called stego-security \cite{Cayre2008}, recalled below.

In the prisoner problem of Simmons~\cite{Simmons83}, Alice and Bob are in jail, and they want to, possibly, devise an escape plan by exchanging hidden messages in innocent-looking cover contents. 
These messages are to be conveyed to one another by a common warden, Eve, who over-drops all contents and can choose to interrupt the communication if they appear to be stego-contents. 
The stego-security, defined in this
framework, is the highest security level in WOA setup~\cite{Cayre2008}.
To recall it, we need the following notations:
\begin{itemize}
 \item $\mathds{K}$ is the set of embedding keys,
 \item $p(X)$ is the probabilistic model of $N_0$ initial host contents,
 \item $p(Y|K_1)$ is the probabilistic model of $N_0$ watermarked contents.
\end{itemize}

Furthermore, it is supposed in this context that each host content has been watermarked with the same secret key $K_1$ and the same embedding function $e$.
It is now possible to define the notion of stego-security.

\begin{definition}[Stego-Security]
\label{Def:Stego-security}
The embedding function $e$ is \emph{stego-secure} if and only if:
$$\forall K_1 \in \mathds{K}, p(Y|K_1)=p(X).$$
\end{definition}

This definition is almost always considered as not really tractable in
practice, reasons explaining this mistrust are outlined in the
following section. 
This is the reason why the information hiding community majorly focuses on
the construction of steganalyzers, supposed to be able to determine
whether a given communication channel appears to transmit steganographied
messages or not. 

\subsection{Drawbacks of the Stego-Security Notion}
\label{sec:drawbacks}

Theoretically speaking, the stego-security notion matches well with the idea of a perfect 
secrecy in the WOA category of attacks. However, its concrete verification raises
several technical problems difficult to get around. These difficulties impact drastically
the effective security of the scheme.

For instance, in a stego-secure scheme, the distribution of the set of watermarked 
images must be the same than the one of the original contents, no matter the chosen keys.
But \emph{how to determine practically the distribution of the original contents?} 
Furthermore, claiming that Alice can constitutes her own subset of well-chosen images having 
the same ``good'' distribution is quite unreasonable in several contexts of 
steganography: Alice has not \emph{always} the choice of the supports. Moreover, 
it introduces a kind of bias, as the warden can find such similarities surprising.
 Suppose however that Alice is in the best situation for her, that is, she has the possibility to 
constitute herself the set of original contents. How can she proceed practically 
to be certain that all media into the set follow a same distribution $p(X)$?
According to the authors opinion, Alice has two possible choices:
\begin{enumerate}
\item Either she constitutes the set by testing, for each new content, whether this media has a same distribution than the ones that have been already selected.
\item Or she forges directly new images by using existing ones. For instance, she can replace all the
least significant bits of the original contents by using a good pseudorandom number generator.
\end{enumerate}

In the first situation, Alice will realize a $\chi^2$ test, or other statistical tests of this kind,
 to determine if the considered image (its least significant bits, or its low frequency coefficients, etc.) has a same distribution than images already selected. In that situation, Alice
 does not have the liberty to choose the distribution, and it seems impossible to find a scheme
being able to preserve any kind of distribution, for all secret keys and all hidden messages.
Furthermore, such statistical hypothesis testing are not ideal ones, as they only regard if a result is unlikely to have occurred by chance alone \emph{according to a pre-determined threshold probability} (the significance level). Errors of the first (false positive) and second kind (false negative) occur necessarily, with a certain probability. In other words, with such an approach, Alice
cannot design a perfect set of cover contents having all the same probability $p(X)$. 
This process leads to a set of media that follows a distribution Alice does not have access to.

The second situation seems more realistic, it will thus be 
further investigated in the next section.

\section{Toward a Cryptographically Secure Hiding}
\label{sec4}
In this section a theoretical framework for information
hiding security is proposed, which is more closely resembling that
of usual approaches in cryptography. It allows to define the 
notion of steganalyzers, it is compatible with the new original
scenarios of information hiding that have been dressed 
in the previous sections, and it does not have the drawbacks
 of the stego-security definition.

\subsection{Introduction}

Almost all branches in cryptology have a complexity approach for security.
For instance, in a cryptographic context, a pseudorandom number generator (PRNG) is a deterministic
algorithm $G$ transforming strings of length $\ell$ into strings of length
$M$, with $M> \ell$. 
The notion of {\it secure} PRNG can be defined as follows~\cite{Yao82}.


\begin{definition}
Let $\mathcal{D}: \mathds{B}^M \longrightarrow \mathds{B}$ be a probabilistic algorithm that runs
in time $T$. 
Let $\varepsilon > 0$. 
$\mathcal{D}$ is called a $(T,\varepsilon)-$distinguishing attack on pseudorandom
generator $G$ if
\begin{small}
$$\left| Pr[\mathcal{D}(G(k)) = 1 \mid k \in_R \{0,1\}^\ell ] - Pr[\mathcal{D}(s) = 1 \mid s \in_R \mathds{B}^M ]\right| \geqslant \varepsilon,$$
\end{small}
where the probability is taken over the internal coin flips of $\mathcal{D}$, and the notation
``$\in_R$'' indicates the process of selecting an element at random and uniformly over the
corresponding set.
\end{definition}

Let us recall that the running time of a probabilistic algorithm is defined to be the
maximum of the expected number of steps needed to produce an output, maximized
over all inputs; the expected number is averaged over all coin flips made by the algorithm~\cite{Knuth97}.
We are now able to recall the notion of cryptographically secure PRNG.

\begin{definition}
A pseudorandom generator is $(T,\varepsilon)-$secure if there exists no $(T,\varepsilon)-$distinguishing attack on this pseudorandom generator.
\end{definition}

Intuitively, it means that no polynomial-time algorithm can make a 
distinction, with a non-negligible probability, between a truly 
random generator and $G$.

Inspired by these kind of definitions, we propose what follows.

\subsection{Definition of a stegosystem}

\begin{definition}[Stegosystem]
Let $\mathcal{S}, \mathcal{M}$, and $\mathcal{K}=\mathds{B}^\ell$ 
three sets of words on $\mathds{B}$ called respectively the sets of 
supports, of messages, and of keys (of size $\ell$).

A \emph{stegosystem} on $(\mathcal{S}, \mathcal{M}, \mathcal{K})$ 
is a tuple $(\mathcal{I},\mathcal{E}, inv)$ such that:
\begin{itemize}
\item $\mathcal{I}$ is a function from $\mathcal{S} \times \mathcal{M} \times \mathcal{K} $ to $ \mathcal{S}$, 
$(s,m,k) \longmapsto \mathcal{I}(s,m,k)=s'$, 
\item $\mathcal{E}$ is a function from $\mathcal{S} \times \mathcal{K}$ to $\mathcal{M}$, 
$(s,k) \longmapsto \mathcal{E}(s,k) = m'$.
\item $inv$ is a function from $\mathcal{K} $ to $\mathcal{K}$, s.t. $\forall k \in
 \mathcal{K}, \forall (s,m)\in \mathcal{S}\times\mathcal{M},$ \linebreak
$ \mathcal{E}(\mathcal{I}(s,m,k),inv(k))=m$.
\item $\mathcal{I}(s,m,k)$ and $\mathcal{E}(c,k^\prime)$ can be computed in
 polynomial time. 
\end{itemize}
 $\mathcal{I}$ is called the insertion or embedding
function, $\mathcal{E}$ the extraction function, $s$ the host content,
$m$ the hidden message, $k$ the embedding key, $k'=inv(k)$ the extraction key, and
$s'$ is the stego-content. If $\forall k \in \mathcal{K}, k=inv(k)$, the stegosystem is symmetric (private-key), otherwise
it is asymmetric (public-key).
\end{definition}

\subsection{Heading Notions}

%
%
%
%

%
%

\begin{definition}[$(T,\varepsilon)-$distinguishing attack]
Let $S=(\mathcal{I},\mathcal{E}, inv)$ 
a stegosystem on $(\mathcal{A}, \mathcal{M}, \mathcal{K})$, with 
$\mathcal{A} \subset \mathds{B}^M$.
A $(T,\varepsilon)-$distinguishing attack on the stegosystem $S$
is a probabilistic 
algorithm $\mathcal{D}:\mathcal{A} \longrightarrow \{0,1\}$ in running time $T$, such that
there exists $m \in \mathcal{M}$, 
\begin{flushleft}
$\left| Pr\left[\mathcal{D}\left(\mathcal{I}(s,m,k)\right)=1 \mid k \in_R \mathcal{K}, s \in_R \mathcal{A}\right]\right.$
\end{flushleft}

\begin{flushright}
 $\left. -Pr\left[\mathcal{D}\left(x\right)=1 \mid x \in_R \mathcal{A}\right]\right|\geqslant \varepsilon$,
\end{flushright}
where the probability is also taken over the internal coin flips of $\mathcal{D}$,
 and the notation $\in_R$ indicates the process of selecting an element at 
random and uniformly over the corresponding set. 
\end{definition}

\begin{definition}
A stegosystem is $(T,\varepsilon)-$un\-dis\-tin\-gui\-sha\-ble if there exists no 
$(T,\varepsilon)-$distinguishing attack on this stegosystem.
\end{definition}

Intuitively, it means that there is no polynomial-time probabilistic algorithm
being able to distinguish the host contents from the stego-contents


\subsection{A Cryptographically Secure Information Hiding Scheme}

\begin{theorem}
 Let $$\mathcal{S} = \left\{s_1^1, s_2^1,\ldots,s_{2^N}^1, s_1^2,
 s_2^2,\ldots,s_{2^N}^2,\ldots,s_1^r, s_2^r,\ldots,s_{2^N}^r\right\}$$
a subset of $\mathds{B}^M = \mathcal{A}.$
Consider $G:\mathds{B}^L \longrightarrow \mathds{B}^N$ a $(T,\varepsilon)-$secure
pseudorandom number generator, and
$\mathcal{I}(s_j^i,m,k)=s^i_{m\oplus G(k)}$. Assuming that $r$ is a
constant, and that from $i,j$ one can
compute the image $s^i_j$ in time $T_1$, the steosystem is
$(T-T_1-N-1,\varepsilon)$-secure. 
\end{theorem}

Intuitively, $\mathcal{S}$ is built from $r$ images containing $N$ bits
of low information. The image $s^i_j$ corresponds to the $i$-th image where
the $N$ bits are set to $j$.

\begin{proof}
Assume there exists a $(T^\prime,\varepsilon)$ distinguisher $\mathcal{D}^\prime$for the
stego-system. Therefore, there exists $m_0$ such that 
 \begin{equation}\label{eq:pch1}
 \begin{array}{l}
 \left| Pr\left(\mathcal{D}^\prime\left(\mathcal{I}(s,m_0,k)\right)=1 \mid k \in_R \mathds{B}^\ell, s \in_R \mathcal{S}\right)\right. \\
 \left. -Pr\left(\mathcal{D}^\prime\left(x\right)=1 \mid x \in_R \mathcal{S}\right)\right|\geqslant \varepsilon
 \end{array}
\end{equation}
Choosing randomly and uniformly $s\in \mathcal{S}$ is equivalent to choose
uniformly and randomly $i\in\{1,\ldots,r\}$ and $j\in\{1,\ldots,2^N\}$.
Therfore (\ref{eq:pch1}) is equivalent to 
 \begin{equation}\label{eq:pch2}
 \begin{array}{l}
 \left| Pr\left(\mathcal{D}^\prime\left(s^i_{m_0\oplus G(k)}\right)=1 \mid k
 \in_R \mathds{B}^\ell, i \in_R \{1,\ldots,r\}\right)\right.\\
 \left.-Pr\left(\mathcal{D}^\prime\left(x\right)=1 \mid x \in_R \mathcal{S}\right)\right|\geqslant \varepsilon
 \end{array}
\end{equation}

Let $\mathcal{D}$ be the distinguisher for $G$ defined for $y\in\{0,1\}^N$ into
$\{0,1\}$ by:
\begin{enumerate}
\item Pick randomly and uniformly $i\in \{1,\ldots,r\}$.
\item Compute $s=s^i_{m_0\oplus y}$.
\item Return $\mathcal{D}^\prime(s)$.
\end{enumerate}
The complexity of this probabilistic algorithm is $1$ for the first step since $r$ is a
constant,
$T_1+N$ for the second step, and $T^\prime$ for the last one. Thus it works in thime 
$T^\prime+T_1+1+N$.

Now we claim that $\mathcal{D}$ is a $(T^\prime+T_1+1+N,\varepsilon)$-distinguisher for $G$. 
Indeed, 
\begin{align*}
& Pr\left(\mathcal{D}\left(y\right)=1 \mid y \in_R \{0,2^N\}\right)\\
=& Pr\left(\mathcal{D}^\prime\left(s^i_y\right)=1 \mid y \in_R
\{0,2^N\},i\in_R\{1,\ldots,r\}\right)\\
=&Pr\left(\mathcal{D}^\prime\left(x\right)=1 \mid x \in_R \mathcal{S}\right).
\end{align*}
Moreover,
\begin{align*}
& Pr\left(\mathcal{D}\left(G(k)\right)=1 \mid k \in_R \{0,1\}^\ell\right)\\
=& Pr\left(\mathcal{D}^\prime\left(s^i_{m_0\oplus G(k)}\right)=1 \mid k
\in_R \{0,1\}^\ell, i\in_R\{1,\ldots,r\}\right).
\end{align*}
Therefore, using (\ref{eq:pch2}), one has
\begin{equation}
\begin{array}{l}
\left| Pr[\mathcal{D}(G(k)) = 1 \mid k \in_R \{0,1\}^\ell ]\right.\\
\left.- Pr[\mathcal{D}(s) = 1 \mid s \in_R \mathds{B}^M ]\right| \geqslant \varepsilon,
\end{array}
\end{equation}
proving that $\mathcal{D}$ is a $(T^\prime+T_1+1+N,\varepsilon)$-distinguisher for
$G$, which concludes the proof. 
\end{proof}

\section{Conclusion}
\label{sec7}

In this research work, a new rigorous approach for 
secure steganography, based on the complexity theory, has
been proposed. This work has been 
inspired by the definitions of security that can usually be found in other
branches of cryptology. 
We have proposed a new understanding for the notion of \emph{secure hiding}
and presented a first secure information hiding scheme. 
The intention was to prove the existence of such a scheme 
and to give a rigorous cryptographical framework for
steganography.

In future work, we will investigate the situation where
detection is impossible. In that case, we will consider 
both weak indistinguability (using a statistical or a
complexity approach, with the cryptographically secure 
definition of PRNGs) and strong indistinguability (using
the well known CC1 and CC2 sets).
Additionally, we will reconsider and improve the definitions
of security in the information hiding literature that are
based on the signal theory. Among other thing, we will 
take into account a Shannon entropy that is not reduced to
simple 1-bit blocs. 
Finally, we will show that tests using generators allow to
attack information hiding schemes that are secure for the
statistical approach, as LSB are not uniform in that situation.

\bibliographystyle{plain}
\bibliography{mabase}

\end{document}